\documentclass{aa}  
\usepackage{graphicx,psfig}
\usepackage{txfonts}
\def \ki2{$\chi^2$}
\def \deltaki2{$\Delta\chi^2$}
\def \hip{{\em Hipparcos}}
\def \t0{T$_0$}
\def \p_hip{$P_{\rm Hipp}$}
\def \tr9{HD\,189733}

\begin{document}
   \title{A posteriori detection of the planetary transit of \tr9\,b
   in the Hipparcos photometry}

   \author{G. H\'ebrard%\inst{1}
           \and
           A. Lecavelier des \'Etangs%\inst{1}
            }

   \offprints{G. H. (\email{hebrard@iap.fr})}

   \institute{Institut d'Astrophysique de Paris, UMR7095 CNRS,
   Universit\'e Pierre \& Marie Curie, 98$^{\rm bis}$ boulevard Arago,
   F-75014 Paris, France} % (\email{hebrard@iap.fr, lecaveli@iap.fr})}

   \date{Received ...; accepted ...}

  \abstract
   {}
   {Thanks to observations performed at the Haute-Provence
   Observatory, Bouchy et al. recently announced the detection of a
   2.2-day orbital period extra-solar planet that transits the disk of
   its parent star, \tr9. We searched in the \hip\ photometry
   Catalogue possible detections of those transits.}
   {Statistic studies were performed on the \hip\ data in order to
   detect transits of \tr9\,b and to quantify the significance of
   their detection.}
   {With high level of confidence, we find that \hip\ likely observed
   one transit of \tr9\,b in October 1991, and possibly two others in
   February 1991 and February 1993. Using the range of possible
   periods for \tr9\,b, we find that the probability that none of
   those events are due to planetary transits but are instead all due
   to artifacts is lower than 0.15\,\%. Thanks to the 15-year temporal
   baseline available, we can measure the orbital period of the planet
   \tr9\,b with a particularly high accuracy. We obtain a period of
   $2.218574^{+0.000006}_{-0.000010}$~days, corresponding to an
   accuracy of $\sim1$~second. Such accurate measurements might provide
   clues for companions presence.}
   {}

   \keywords{Stars: individual: \tr9 -- planetary systems}

\authorrunning{H\'ebrard \& Lecavelier}
\titlerunning{A posteriori detection of the planetary transit of 
\tr9\,b with Hipparcos}
%\titlerunning{A posteriori detection of \tr9\,b with Hipparcos}

   \maketitle
%
%________________________________________________________________

\section{Introduction}

Bouchy et al.~(\cite{bouchy05}) recently announced the detection of a
2.2-day orbital period extra-solar planet that transits the disk of
its parent star, the dwarf \tr9, which is located only
10~arcmin from the famous Dumb-Bell Nebula. This detection was
performed thanks to spectroscopic and photometric data collected at
the Haute-Provence Observatory, France, as part of the ELODIE
metallicity-biased search for transiting hot Jupiters (Da~Silva et
al.~\cite{dasilva05}).  Together with radial velocity measurements,
observations of transits allow the actual mass and radius of an
extra-solar planet to be measured. Transiting planets also allow
follow-up observations to be performed during transits (Charbonneau et
al.~\cite{charbonneau02}; Vidal-Madjar et al.~\cite{avm03},
\cite{avm04}) or anti-transits (Charbonneau et
al.~\cite{charbonneau05}; Deming et al.~\cite{deming05}), yielding
physical constraints on the atmospheres of these planets.

To date, \tr9\,b is only the ninth known transiting
extra-solar planet (Bouchy et al.~\cite{bouchy05}), and the third
transiting a star bright enough to be in the \hip\ Catalogue (Perryman
et al.~\cite{perryman97}). The Epoch Photometry Annex of the \hip\ 
Catalogue contains between $\sim40$ and $\sim300$ measurements
performed during the 1990-1993 mission for each of the 118\,204 stars
of the Catalog. Transits of HD\,209458\,b, the first known
transiting extra-solar planet (Charbonneau et
al.~\cite{charbonneau00}, Henry et al.~\cite{henry00}, Mazeh et
al.~\cite{mazeh00}), were a posteriori detected in \hip\ data by
Robichon \& Arenou~(\cite{robichon00}), Castellano et
al.~(\cite{castellano00}), and S\"oderhjelm~(\cite{soder99}). The
transits of HD\,149026\,b (Sato et al.~\cite{sato05}) are not
deep enough (0.003~mag) to be detectable with \hip\ (H\'ebrard et
al.~\cite{hebrard05}).  Here we show that transits of \tr9\,b were
detected by \hip, and we quantify the signification of this a
posteriori detection.  The long available temporal baseline allows us
to obtain an accurate orbiting period of this hot Jupiter.

\section{Folding the \hip\ photometric measurements}
\label{sect_folding}

\subsection{\hip\ photometric data of \tr9}
\label{subsect_data}

\begin{figure}[h]
\begin{center}
\psfig{file=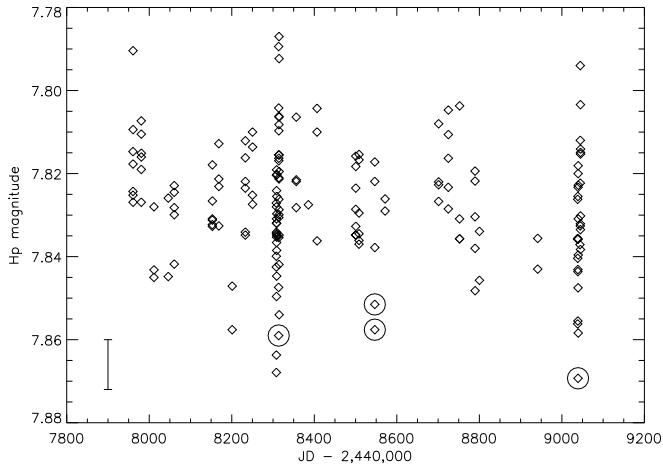,height=6.5cm} \caption{The 176 reliable
\hip\ photometric measurements of \tr9. The four measurements
performed during planetary transits of \tr9\ are surrounded by
circles. The typical error bar on $H_p$ is represented on the
bottom, left. \label{fig_obs} }
\end{center}
\vspace{-0.7cm}
\end{figure}

The \hip\ Catalogue includes \tr9\ photometry measurements at 185
different epochs. We only used in the present study the 176
measurements that are ``accepted'' in the Catalogue; the 9 remaining
ones are flagged in the Catalogue as perturbed and not reliable.
These 176 values are plotted in Fig.~\ref{fig_obs} over the 3-year
observation baseline. The epochs of the measurements are given in
Terrestrial Time corresponding to the Solar System Barycentric Julian
Date (BJD). The differences between BJD and Heliocentric Julian Date
(HJD) is negligible for our study. The \hip\ measurements were
performed in the specific $H_p$ band, which is centered near
4500\,\AA\ and has a width of $\sim 2400$\,\AA. The estimated standard
errors of each individual $H_p$ magnitude are around 0.012~mag; this
makes the $\sim3$\,\% deep \tr9\ planetary transit in
principle~detectable.

Two sets of numerous, dispersed measurements performed at two
neighboring epochs are apparent in Fig.~\ref{fig_obs}. Owing to the
\hip\ scanning law, there are actually four time intervals 
of about 1.5~day each (BJD\,--\,2,440,000\,=\,8308.5, 8314.3, 9039.0,
and 9044.7) during which numerous photometric measurements were
performed.  The dispersion of these measurements shows the stellar
variability of \tr9, which is classed as microvariable in the \hip\ 
Catalogue. As we see below, the microvariability of \tr9\ does not
prohibit transits detection (see also
\S~\ref{subsect_microvariability}).

\subsection{Period search}
\label{subsect_search}

\tr9\,b orbits its parent star every $\sim2.2$~days with a transit
duration of $\sim1.6$~hour, so about 3\,\% of randomly chosen
observation would be expected to fall during the transit. This
corresponds to about 5 measurements in the case of the 176
available \hip\ values, which are however not regularly sampled in
time as we described above. Nevertheless, it is likely that a
few planetary transits were sampled in these 176 measurements.

We performed a \ki2\ analysis in order to attempt to detect transits.
We scanned the possible periods around the period of 2.2190~days given
by Bouchy et al.~(\cite{bouchy05}) with steps of $5\times10^{-7}$~day
(or about 0.04 sec), in the range [$2.217-2.221$]~days, that is four
times the uncertainties given by Bouchy et al.~(\cite{bouchy05}).
Note that a broader search was also performed (see
\S~\ref{subsect_large_period}). The phase of the transit within the
\hip\ data is a function of the assumed period. Indeed, for a given
period, the phase is strongly constrained by the mean transit epoch,
\t0, as determined by the ground-based discovery and follow-up
observations. As, for a given period, there are integer numbers of
\tr9\,b orbits between the transits observed by Bouchy et
al.~(\cite{bouchy05}) and the ones possibly detected by \hip, the
accuracy of the phase is exactly the accuracy of \t0. Bouchy et
al.~(\cite{bouchy05}) reported
\t0$\,=2,453,629.3890\pm0.0004$ (HJD) so the uncertainty on
the phase in the \hip\ data is $0.000018$, corresponding to
0.0004~day. This assumes that the period is constant, or at least that
if any, the variations of the period are small, with a constant
average value.

For each of the 8000 periods tested, we computed the \ki2, i.e. the
quadratic sum of the weighted difference between the observed
magnitudes and a transit model.  The transit model is an approximation
of the light curve presented by Bouchy et al.~(\cite{bouchy05}). It
assumes a 2.7\,\% deep transit, and durations from the 1$^{\rm st}$ to
the 4$^{\rm th}$ contacts and from the 2$^{\rm nd}$ to the 3$^{\rm
rd}$ contacts of 1.60~hour and 0.66~hour,~respectively.

\begin{figure}[h]
\begin{center}
\psfig{file=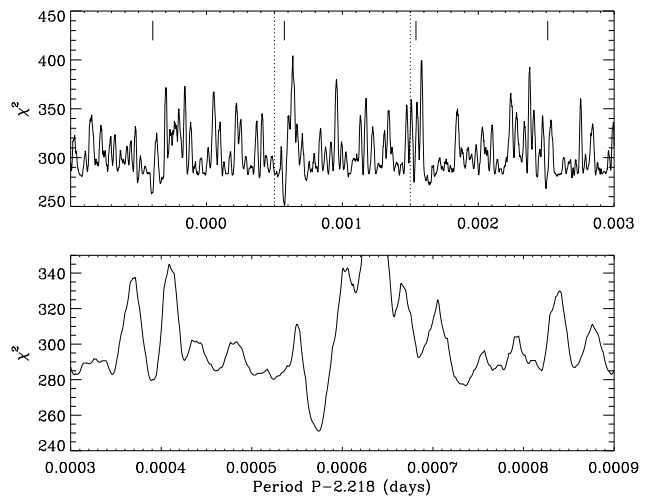,height=6.5cm,width=8.5cm,angle=90}
\caption{{\it Top}: \ki2\ of the fit of the \hip\ magnitudes with
a transit light curve, as a function of the trial period. The period
interval allowed by Bouchy et al.~(\cite{bouchy05}) is represented by
vertical dotted lines.  The ticks show the positions of the possible
periods assuming that a transit occurred on 1991, Oct. 17$^{\rm th}$
(see \S~\ref{subsect_october_1991}).  {\it Bottom}: Zoom around the
minimum \ki2, found for a 2.218574-day~period. \label{fig_chi2} }
\end{center}
\vspace{-0.7cm}
\end{figure}

\begin{figure}[h]
\begin{center}
\psfig{file=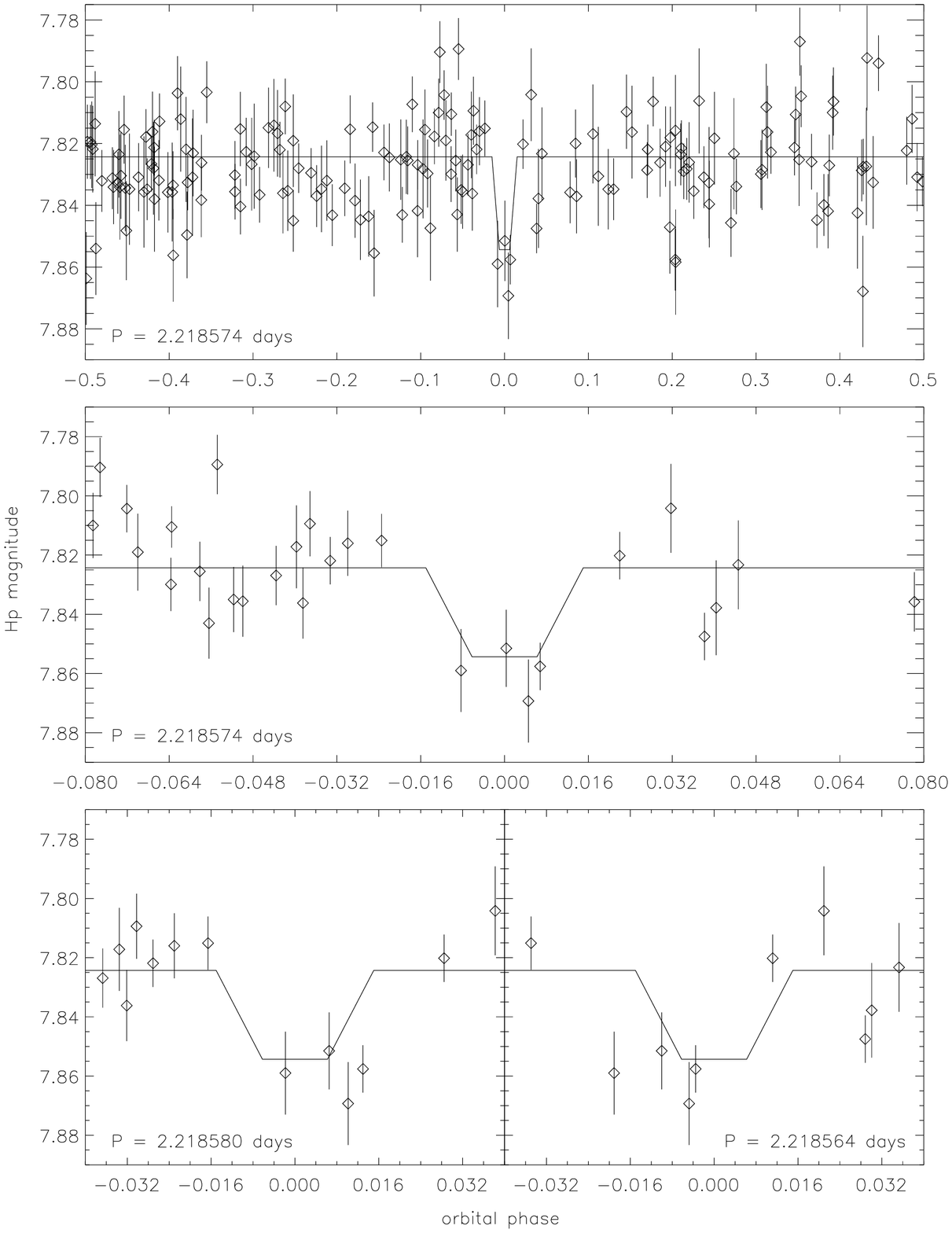,height=11cm} \vspace{0.5cm} \caption{{\it
Top}: \hip\ photometric measurements folded with a period
\p_hip$\,=2.218574$~days. We found this period from \ki2\ analysis
of the \hip\ measurements. It agrees with the value reported  by
Bouchy et al.~(\cite{bouchy05}). The approximation of the transit
curve from Bouchy et al.~(\cite{bouchy05}) used for the \ki2\
computation is over-plotted. {\it Middle}: Zoom on the measurements
around the transit (phase = 0). {\it Bottom}: Same plots, but for
the two extreme values of the error bar on \p_hip\ (see
Sect.~\ref{sect_accuracy}). \label{fig_fold} }
\end{center}
\vspace{-0.7cm}
\end{figure}

Fig.~\ref{fig_chi2} shows the \ki2\ as a function of the trial period,
which is the only free parameter. A clear minimum is seen for the
period \p_hip~=~2.218574~days. The minimum is \ki2$\,=251.0$. We
attribute this high \ki2, considering the 175 degrees of freedom, to
the microvariability of \tr9\ (\S~\ref{subsect_microvariability}).
The value of the orbital period which we found, \p_hip, is in
agreement with the one reported by Bouchy et al.~(\cite{bouchy05}),
$2.2190\pm0.0005$~days. The \hip\ data folded with the period \p_hip\
are plotted in Fig.~\ref{fig_fold}.

The median value for $H_p$ is 7.827. Assuming that \p_hip\ is the
orbital period of \tr9\,b, the weighted average value of the four
$H_p$ measurements obtained during the planetary transit is
$7.859\pm0.006$ whereas the weighted average of the 172 remaining
points is $7.8261\pm0.0009$.  Thus, the depth of the transit light
curve as measured with \hip\ is ($0.033\pm0.006$)~mag or
($3.1\pm0.6$)\,\% in flux.  This agrees with the light curve reported
by Bouchy et al.~(\cite{bouchy05}) from accurate and well sampled
ground-based observations of the transit. One can note that the
\hip\ data of HD\,209458, the host of the first known transiting
planet (Charbonneau et al.~\cite{charbonneau00}, Henry et
al.~\cite{henry00}), yield a transit marginally deeper than the actual
one, which favored its detection in the \hip\ photometry (Robichon \&
Arenou~\cite{robichon00}, Castellano et al.~\cite{castellano00}). This
is not the case for \tr9.

\subsection{Transits observations}
\label{subsect_transits}

\begin{figure*}
\begin{center}
\psfig{file=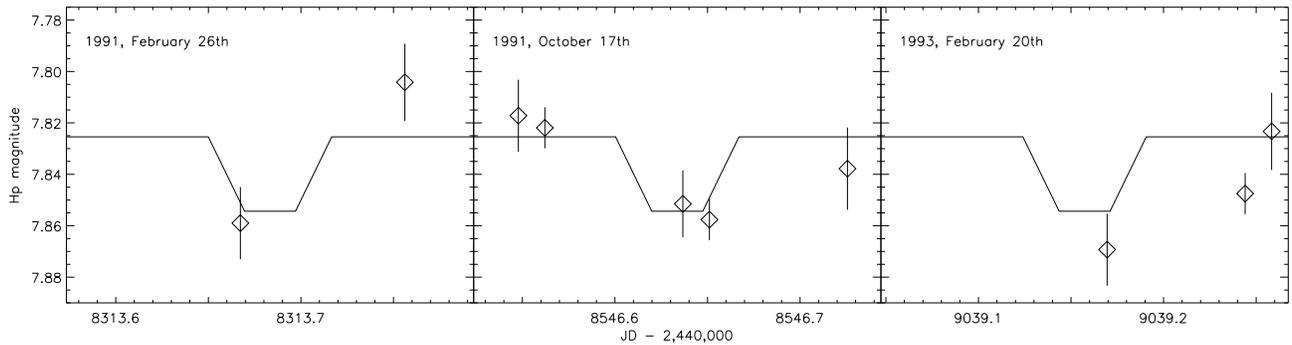,height=4.cm} \vspace{0.5cm}
\vspace{0.5cm}
\caption{Photometry around the epochs of the three planetary
transits of \tr9\ observed by \hip. The model of the transit
assuming a planetary orbital period \p_hip$\,=2.218574$~days is
over-plotted.
\label{fig_3transits}
}
\end{center}
\vspace{-0.7cm}
\end{figure*}

Assuming \p_hip, four \hip\ photometric measurements are clearly
located within the planetary transit (see Fig.~\ref{fig_fold}).  They
sample actually three different transits, which are surrounded by
circles in Fig.~\ref{fig_obs}. The transits occurred on 1991,
Feb. 26$^{\rm th}$ (BJD 2,448,313.68), 1991, Oct.  17$^{\rm th}$ (BJD
2,448,546.63), and 1993, Feb. 20$^{\rm th}$ (BJD 2,449,039.16). The
\hip\ measurements performed around these three dates are plotted in
Fig.~\ref{fig_3transits}. Between these three transits observed by
\hip\ and this observed by Bouchy et al.~(\cite{bouchy05}) on 2005,
Sept. 15$^{\rm th}$ at the Haute-Provence Observatory, there were
exactly 2396, 2291, and 2069 orbits of the planet \tr9\,b around its
host star.

\section{Significance of the detection}
\label{sect_significance}

\hip\ measurements have a poor time coverage. In addition, the
errors on each measurement are of the same order of magnitude than
the expected transit effect. Moreover, \tr9\ seems to exhibit an
actual stellar microvariability (see \S~\ref{subsect_data}). 
Due to these causes, various periods might be found such that the
\tr9\ \hip\ photometric measurements are consistent with a
transit light curve and agree with the period and \t0\ given by Bouchy
et al.~(\cite{bouchy05}). The question is to know whether the period
\p_hip\ we report above produces or not a solution that is
significantly better than those obtained with other periods. Three
arguments allow us to answer this question in the affirmative: \ki2\ 
variations, fits with an inverse light curve, and a bootstrap test.

\begin{figure}[h]
\begin{center}
\psfig{file=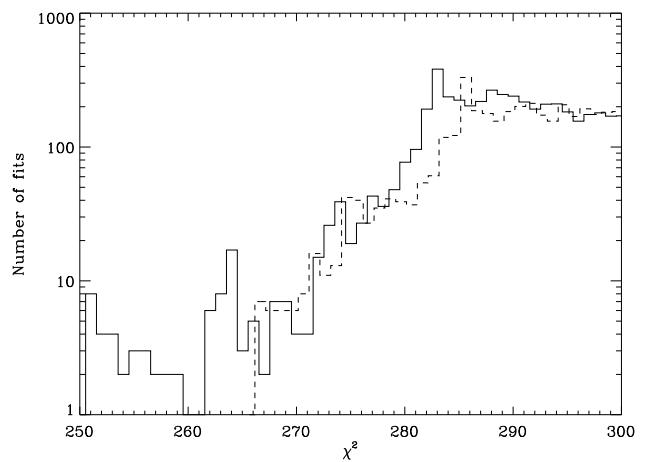,height=6.5cm,angle=90}
\caption{\ki2\ histogram (solid line) of the fits performed with
8000 different planetary orbital periods chosen in the range
[$2.217-2.221$]~days. The solution with \p_hip\ clearly emerge from
the \ki2\ distribution, at \ki2$=251.0$. The dotted line shows the
corresponding histogram in the case of fits with inverse light curve;
no solutions emerge in that case, where no signal is present (see
Sect.~\ref{sect_significance}). \label{fig_histo_chi2} }
\end{center}
\vspace{-0.7cm}
\end{figure}

First, as seen on Fig.~\ref{fig_chi2}, the \ki2\ of the solution with
\p_hip\ is significantly lower than those obtained with other
periods. Some \ki2\ local minima are found for other periods in the
range [$2.217-2.221$]~days; however, the lowest ones present a
\ki2\ at least greater by $\sim11$ from the minimum \ki2\ found for
\p_hip. A \deltaki2\ of 11 is significant. This is seen in 
Fig.~\ref{fig_histo_chi2}, that plots the \ki2\ histogram of the fits
performed with the 8000 different planetary orbital periods chosen in
the range [$2.217-2.221$]~days (see \S~\ref{subsect_search}). As
expected if no signal is present, there is a continuous distribution
of \ki2, with fewer solutions with lower \ki2; this is seen in the
tail of the \ki2\ distribution.  However, there is a solution that
emerges from this distribution, at \deltaki2~$\simeq11$ from the end
of the distribution tail, namely \p_hip. This detection is
thus~significant.

Second, we performed the same \ki2\ scan as that presented in
\S~\ref{subsect_search}, but with an inverse model light curve,
i.e. an increase of the star brightness of 2.7\,\% instead of a
decrease. If the signal we report is just an artifact due to the noise
present in the data and the high number of folding possibilities, the
chances to find a false absorption light curve should be roughly the
same than those to find a false emission light curve. However, the
lowest \ki2\ found in the case of these inverse light curves are
around 266.7, which is larger by at least \deltaki2~$\simeq15.7$ from
the lowest \ki2 reported in \S~\ref{subsect_search} for the normal
light curve with \p_hip.  The \ki2\ histogram performed in the case of
the inverse light curve shows a decreasing tail toward lower \ki2\
values, but without any solution emerging from this tail (see
Fig.~\ref{fig_histo_chi2}).  Thus, inverse light curves do not show
any significant solution within the period range defined by Bouchy et
al.~(\cite{bouchy05}).

Finally, we performed a bootstrap experiment to quantify the
significance of our detection and the probability that the three
transits apparently detected in the period range [$2.217-2.221$]~days
can be all due to noise.  We generated 20\,000 random sets of data
from the original \hip\ data by redistribution of the times of
observation (we kept the times of observation the same, but
scrambled the photometric values). This method includes all source of
noise in the real data, including the observed microvariability of the
star. From these 20\,000 trials, only 30 give a detection of a period
in the range of acceptable values and a lower \ki2. With a false-alarm
probability of less than 0.15\,\%, this gives us confidence that the
transits detection we report in Sect.~\ref{sect_folding} in the \hip\ 
data is real. Note that if we restrain the period search to the exact
range allowed by Bouchy et al.~(\cite{bouchy05}), the false-alarm
probability decreases to 7/20\,000, or 0.035\,\%.

The false-alarm probability is actually even lower. Indeed, as these
four points appear to be real transit measurements, they will favor
false solutions (i.e. with other periods) to be found in the bootstrap
test since these four low points allow transits curves to be~fitted.

Thus, we conclude that the orbital period presented in
Sect.~\ref{sect_folding} results from an actual detection with
\hip\ of transits of \tr9\,b in front of its parent star.

\section{Accuracy of the planetary orbital period}
\label{sect_accuracy}

We quantify here the accuracy of the \tr9\,b orbital period we
obtained, which is computed from \ki2\ variations. \deltaki2$\,=11$
appears to be a reasonable confidence interval, as the local
minima reported in Sect.~\ref{sect_significance} have at least this
\deltaki2\ with our best solution. According to Fig.~\ref{fig_chi2}, an
interval with \deltaki2$\,=11$ implies an error bar on \p_hip\ of
$^{+0.000006}_{-0.000010}$~day, corresponding to about
$^{+0.5}_{-0.9}$~second. The \hip\ data folded with the two extreme
periods of this interval are plotted on the lower panel of
Fig.~\ref{fig_fold}. Thanks to the 15-year baseline, this error bar is
almost 100 times smaller than that obtain by Bouchy et
al.~(\cite{bouchy05}) on a one-week baseline.

There are two other causes of uncertainties on \p_hip\ that are not
included in this \ki2\ study. However, they are negligible. The first
one is due to the uncertainty in the mean transit epoch \t0.  If \t0\
is sooner or later, the obtained period would be respectively shorter
or longer. This error on \p_hip\ is equal to the uncertainty on \t0\
divided by the largest number of planetary orbits between two observed
transits, namely 2396 (see \S~\ref{subsect_transits}). Bouchy et
al.~(\cite{bouchy05}) reported a 0.0004-day error in \t0, which thus
translates into an error of $1.7\times10^{-7}$~day on \p_hip; this is
about 40 times lower than the error bar reported above.

The second extra uncertainty, which is due to the shape and the
duration of the transit, is even lower, as those parameters are well
known from the photometric observations of Bouchy et
al.~(\cite{bouchy05}). We note that fitting the transit with a
box-shape approximation might lead to erroneous solution as the impact
parameter of \tr9\,b is relatively high. This has however no effect on
our solution, as the four points we identified in the transits are
located in the central part of the transit, and none is located near
the beginning of the end of them.

Finally, we also performed all the tests described in the present
paper using the detection statistic $l$ as described by Castellano et
al.~(\cite{castellano00}).  All the results in terms of period
determination, error bars, and significance of the detection are
identical to those obtained using the \ki2.

\section{Discussion}

\subsection{Periods from the October 1991's transit}
\label{subsect_october_1991}

The period determination and the deep minimum of the \ki2\ are mainly
based on the detection of a transit on 1991, October 17$^{\rm th}$.
In that case, two measurements were obtained just before the transit,
two other during it, and a last one just after the transit
(Fig.~\ref{fig_3transits}, middle panel).  The observations of the two
other transits have only one point during the transit and lower
quality flags (\S~\ref{subsect_quality_flags}).  However, using this
single transit of October 1991, an accurate period can also be
estimated.  If $n$ is the number of periods between this transit and
the transit observed by Bouchy et al.~(\cite{bouchy05}) on 2005,
Sept. 15$^{\rm th}$, the period must be $P_n=5082.75/n\pm 9\times
10^{-6}$~days. These possible periods are represented by ticks
on Fig.~\ref{fig_chi2}, upper panel (for $n=\,$2290, 2291, 2292,
and 2293).  It is noteworthy that the best period
\p_hip$\,=P_{n=2291}$ given the lowest \ki2$\,=251.0$ is found in the
Bouchy et al.~(\cite{bouchy05}) range of possible periods. For other
values of $n$ around 2291, we find significantly higher \ki2, in most
cases because other data points are obviously incompatible with the
observation of a transit if they are folded with the corresponding
periods. We note that $P_{n=2287}$, $P_{n=2292}$, and $P_{n=2293}$
(respectively 2.22245, 2.21760, and 2.21664~days) give low
values of \ki2\ (but still higher than 263).  These periods correspond
to the situation where the only transit observed by
\hip\ would be the one of 1991, October 17$^{\rm th}$; they can be 
eliminated only because they are beyond the error bars given by Bouchy
et al.~(\cite{bouchy05}).  The period $P_{n=2294}$ corresponds to the
situation where \hip\ would have observed during two other transits;
for the first one, two measurements are consistent with the transit
light curve, while for the second one, one low quality flagged
measurement is not consistent with the transit light curve.  Again,
this period can be eliminated only because it is well beyond the error
bars given by Bouchy et al.~(\cite{bouchy05}).

\subsection{Large period range scan}
\label{subsect_large_period}

In \S~\ref{subsect_search}, we found the best period by searching for
the period giving a relatively deep minimum for the \ki2\ over four
times the period range given by Bouchy et al.~(\cite{bouchy05}). To
check if such a deep minimum is frequent with the actual \hip\
measurements, we extend the period range and find that such minimum
can also be found if we consider periods down to 2.0822~days or up to
3.4909~days. These periods are far from the Bouchy et
al.~(\cite{bouchy05}) acceptable values, by more than 100~times their
error bar. This strengthens the case that the 2.518574-day period is
peculiar and not simply the best period among statistical variations
of the \ki2.

We note in addition that this 2.0822-day period has a deep
minimum in \ki2\ mainly because of the data of 1991, Oct. 17$^{\rm
th}$. Therefore, it corresponds to folding the data with $n=2441$
orbital periods between this transit observations and the transit
observations at Haute-Provence Observatory performed by Bouchy et
al.~(\cite{bouchy05}) at \t0$\,=2,453,629.3890$. This shows that it is
unlikely that the data can be folded by a period inside the limited
period range of Bouchy et al.~(\cite{bouchy05}) to fit the transit
light curve only because of the statistical noise; this gives us
confidence that we really detected \tr9\,b transits.

\subsection{Stellar microvariability}
\label{subsect_microvariability}

\tr9\ is known to be microvariable. We therefore have to address if 
this variability can mimic a transit light curve. We performed tests
on the long term and short term variability which indicate that this
variability is unlikely to reproduce the observed transit signature.

First, on the long term, we searched for periodicity and found that
\tr9\ presents significant periods of 13.3, 11.8, 8.8, and 4.6~days 
(for the method, see Lecavelier des \'Etangs et
al.~\cite{lecavelier05}).  By removing these periods, we found that
the \ki2\ is significantly reduced to \ki2$\,=193.2$ if we fit the
data with a sinusoid and a period of 11.8~days.  This period is
similar to the stellar rotational period of $\sim11$~days reported by
Bouchy et al.~(\cite{bouchy05}).  This confirms that the large \ki2\
for 175 degrees of freedom is effectively due to the variability of
the star (\S~\ref{subsect_search}).  However, using the data corrected
for these periodic variations, we do not find any significant change
in the period of the planet, its uncertainties, and the significance
of the detection.

It is also desirable to estimate the risk that short term
microvariability (on hours time scale) can mimic the transit light
curve. The bootstrap test presented in Sect.~\ref{sect_significance}
shows that short-term stellar variations are extremely unlikely to be
responsible for our signal. However, this test assumes that there are
no correlations between the different measurements, which can be
incorrect in case of stellar variations.  We performed two different
extra tests in order to take correlations into account.

First, we searched for the couples of measurements that are separated
by less than 0.66~hour and estimated their difference with the mean
brightness of the star. We found 77 such couples and among them only
two couples show differences above 0.025~mag, including the
October 1991 couple of measurements for which the difference is
believed to be due to a real transit. If we correct for the 11.8~day
periodic variations, the October 1991 couple of measurements is the
only one presenting a difference with the mean brightness above
0.02~mag.  This demonstrates that the microvariability is
unlikely to produce in short time two subsequent measurements
reproducing an apparent light decrease similar to the transit
signature.  

As a second test, we search for the groups of four measurements that
are separated by less that 5~hours and estimated the difference
between the mean of the two first measurements and the mean of the two
last.  We found 74 such groups and among them only three groups show
differences above 0.025~mag, including two groups of measurements for
which the difference is believed to be due to a real transit (October
1991 and February 1993).  This again demonstrates that the
microvariability is unlikely to produce in short time four subsequent
measurements reproducing an apparent light decrease similar to the
transit signature of October 1991.

\subsection{Quality flags}
\label{subsect_quality_flags}

The analysis we report above was performed on the 176 \tr9\
measurements ``accepted'' in the \hip\ Catalogue. This includes
the values with the Quality flags ``0'', ``1'', and ``2''; the 9
remaining \tr9\ \hip\ measurements have larger Quality flags,
meaning there are perturbed and unreliable. 17 of the 176 reliable
measurements are flagged ``1'' or ``2'', which means that one of
the two consortia that reduced the data, namely NDAC and FAST,
rejected it. Two of the four measurements located within transits
(those of February 1991 and February 1993) present such flags.
This makes them possibly unreliable. We thus performed all the
tests described above using only the 159 \hip\ measurements of
\tr9\ that are are flagged ``0''. We found the same value for
\p_hip\ within the period range allowing by Bouchy et
al.~(\cite{bouchy05}), but of course, only one transit was
detected (that of October 1991). This makes us confident that our
result is not due to unreliable points.

Interesting enough, a large scan with only these 159 points allow
another period to be found, far from the Bouchy et
al.~(\cite{bouchy05}) range, namely 2.217675~days. This solution
presents a lower \ki2\ than the solution at 2.218574~days. Two
transits are detected in that case, that of October 1991, and
apparently another on 1990, November 5$^{\rm th}$.  The bootstrap test
similar to that presented in Sect.~\ref{sect_significance} indicates
that there is less than 1.5\,\% probability that no transits are
detected in that case, and that this solution is only due to noise.
As this solution is not allowed by Bouchy et al.~(\cite{bouchy05}), it
can not be adopted, except if a second planet is present in the
system, implying a smooth oscillation of the observed transit period
(indeed, radial velocity measurements show motions of the star around
the center of mass of the whole system, whereas transits show motion
of the planet with respect to the central star only). This seems
however unlikely to us. More probably, the low brightness observed on
1990, November 5$^{\rm th}$ is due to the stellar microvariability.
Indeed, these low points are not low anymore if the 11.8-day stellar
oscillations is removed (see \S~\ref{subsect_microvariability}).  This
second bootstrap test reinforces the significance of the \hip\ 
detection of the October 1991 \tr9\,b transit.

\section{Conclusion}

We report the a posteriori detection with \hip\ of three transits
of \tr9\,b in front of its parent star. This allows an accurate
orbital period of this extra-solar planet to be measured.

One valuable question is to know whether an a priori detection of
\tr9\,b would have been possible. Searching planetary candidates in the
\hip\ data seems difficult due to the poor time coverage and the
accuracy of the photometry. Laughlin~(\cite{laughlin00}) and Jenkins
et al.~(\cite{jenkins02}) concluded that \hip\ Catalog does not
represent a likely place to detect planets in the absence of other
information, even if it might provide planetary transit candidates for
follow-up observations. H\'ebrard et al.~(\cite{hebrard05}) made
radial-velocity measurements on transiting candidates selected in
\hip\ Photometry Annex but did not report any detection.

The identification of \tr9\,b transits within \hip\ data was not
obvious. However, the allowed period range was small, the mean time of
the transits as well as their shape and duration were well known,
reducing the number of possible solutions. Searching planetary
transits without a priori knowledge of a period estimation would
require to explore a period range of several days, with numerous,
small steps. As the time of the potential transits are not known, all
the phases should be explored, here again with a small enough
step. Finally, since the impact parameter is not known, as well of the
stellar size relative to the planetary companion nor the limb
darketing, several shapes and durations of the transits should be
looking for. This makes huge the number of solutions in folding the
data.  According the poor time coverage and the accuracy of the
photometry, the \hip\ Photometry Annex does not look to us as a
promising and efficient tool for transits searches without a priori
information. However, it probably will be a valuable database for the
studies of the forthcoming transits that should be discovered in the
future. It would allow them to be quickly confirmed, and their period
to be accurately determined, which is useful for follow-up
observations. It may also reveal long-term period oscillations,
yielding clues for companions presence.

\ 

Shortly before the submission of this paper, we became aware that
Bouchy et al.~(\cite{bouchy05}) also report detection of \tr9\,b
within the \hip\ data. Although their reported error bar on the period
is smaller than ours, their folding solution and period are in good
agreement with ours, which confirms our results.

\begin{acknowledgements}
We thank Fran\c cois Bouchy and the ELODIE Exoplanets Team to have
advised us of their discovery of \tr9\,b before its publication, as
well as~No\"el~Robichon, Fr\'ed\'eric Pont, and David Ehrenreich for
useful discussions.
\end{acknowledgements}

\end{document}